# Imaging moiré flat bands in 3D reconstructed WSe$_2$/WS$_2$ superlattices

*Hongyuan Li*[1,2,3,8], *Shaowei Li*[1,3,4,8]*, *Mit H. Naik*[1,3,8], *Jingxu Xie*[1], *Xinyu Li*[1], *Jiayin Wang*[1], *Emma Regan*[1,2,3], *Danqing Wang*[1,2], *Wenyu Zhao*[1], *Sihan Zhao*[1], *Salman Kahn*[1], *Kentaro Yumigeta*[5], *Mark Blei*[5], *Takashi Taniguchi*[6], *Kenji Watanabe*[7], *Sefaattin Tongay*[5], *Alex Zettl*[1,3,4], *Steven G. Louie*[1,3]*, *Feng Wang*[1,3,4]*, *Michael F. Crommie*[1,3,4]*

[1]Department of Physics, University of California at Berkeley, Berkeley, CA, USA.

[2]Graduate Group in Applied Science and Technology, University of California at Berkeley, Berkeley, CA, USA.

[3]Materials Sciences Division, Lawrence Berkeley National Laboratory, Berkeley, CA, USA.

[4]Kavli Energy Nano Sciences Institute at the University of California Berkeley and the Lawrence Berkeley National Laboratory, Berkeley, CA, USA.

[5]School for Engineering of Matter, Transport and Energy, Arizona State University, Tempe, AZ, USA.

[6]International Center for Materials Nanoarchitectonics, National Institute for Materials Science, Tsukuba, Japan

[7]Research Center for Functional Materials, National Institute for Materials Science, Tsukuba, Japan

[8]These authors contributed equally: Hongyuan Li, Shaowei Li and Mit H. Naik

**Abstract:**

Moiré superlattices in transition metal dichalcogenide (TMD) heterostructures can host novel correlated quantum phenomena due to the interplay of narrow moiré flat bands and strong, long-range Coulomb interactions[1-5]. However, microscopic knowledge of the atomically-reconstructed moiré superlattice and resulting flat bands is still lacking, which is critical for fundamental understanding and control of the correlated moiré phenomena. Here we quantitatively study the moiré flat bands in three-dimensional (3D) reconstructed $WSe_2/WS_2$ moiré superlattices by comparing scanning tunneling spectroscopy (STS) of high quality exfoliated TMD heterostructure devices with *ab initio* simulations of TMD moiré superlattices. A strong 3D buckling reconstruction accompanied by large in-plane strain redistribution is identified in our $WSe_2/WS_2$ moiré heterostructures. STS imaging demonstrates that this results in a remarkably narrow and highly localized K-point moiré flat band at the valence band edge of the heterostructure. A series of moiré flat bands are observed at different energies that exhibit varying degrees of localization. Our observations contradict previous simplified theoretical models but agree quantitatively with *ab initio* simulations that fully capture the 3D structural reconstruction. Here the strain redistribution and 3D buckling dominate the effective moiré potential and result in moiré flat bands at the Brillouin zone K points.

Moiré superlattices in two-dimensional (2D) heterostructures provide an attractive platform to explore novel correlated physics since nearly flat electronic bands can be engineered to enhance the effects of electron-electron correlations. This was first seen in graphene-based moiré superlattices where correlated insulator states, superconductivity, and ferromagnetic Chern insulators have been observed in both twisted bilayer[6-8] and ABC trilayer[9-11] moiré system. The TMD-based moiré superlattice can have even flatter minibands, thus enhancing the role of the long-range Coulomb interactions. They have recently emerged as a complementary model system to explore novel correlated quantum phenomena, such as the correlated insulators and generalized Wigner crystal[1, 4, 5]. More exotic emerging states, such as charge transfer insulators and pair density waves, have been predicted to emerge from theoretical models of hole-doped $WSe_2/WS_2$ moiré heterostructures[12, 13]. These correlated phenomena, however, depend sensitively on the precise structural and electronic properties of the underlying moiré superlattice due to the delicate interplay among atomic geometry, moiré band structure and Coulomb interactions. Fundamental understanding and quantum control of TMD-based moiré phenomena thus require both quantitative knowledge of 3D superlattice reconstructions at the atomic level and flat band electronic structure at the meV energy level, something that has hitherto been missing.

Scanning tunneling microscopy (STM) provides a powerful tool to characterize the atomic and electronic structure of moiré superlattices. Previous STM studies have successfully observed localized moiré flat bands and correlated electronic gaps in twisted bilayer graphene[14-17], and demonstrated moiré site-dependent electronic structure in TMD moire superlattice[18-20].

Narrow moiré flat bands at the valence band edge of TMD heterostructure, however, have not yet been reported. Part of the challenge is the difficulty in fabricating high-quality exfoliated TMD moiré heterostructures on insulating substrates that are suitable for STM characterization. As a result, previous STM studies often focused on chemical vapor deposition (CVD) grown TMD heterostructures on conducting graphite. CVD growth, however, yields lower sample quality and very limited control of stacking order and twist angle of the TMD heterostructure compared to exfoliation and stacking techniques. A graphite substrate can also pin the Fermi level of the TMD material and induce undesirable electronic screening and modification of TMD band structure[21-23].

In this work we determine the moiré flat band electronic structure of three-dimensional (3D) reconstructed $WSe_2/WS_2$ moiré superlattices by combining scanning tunneling spectroscopy (STS) of high quality exfoliated TMD heterostructure devices with *ab initio* simulations of both atomic geometry and electronic band structure of TMD moiré superlattices. Our STM imaging and theoretical simulations reveal a striking 3D buckling reconstruction of the $WS_2/WSe_2$ heterostructures that is accompanied by strong strain redistribution within the moiré superlattice. We observe multiple moiré flat bands at the valence band edge that originate from the K-point, as well as a separate set of deep lying moiré flat bands that originate from the Γ-point (Our convention is to refer the K- and Γ-points of the unfolded $WSe_2$ Brillouin zone (BZ) instead of the moiré BZ). The top-most valence flat band from K-point is prominently narrow with a width of only 10 meV, and is expected to be responsible for recently observed novel correlated insulator behavior and generalized Wigner crystal states[1, 3-5]. The strong localization of this band at the $B^{Se/W}$ stacking site revealed by STS spatial mapping contradicts previous simplified density functional theory (DFT) calculations which predict localization at the AA site[12, 13, 18, 19]. The STS results, however, are fully consistent with our DFT results obtained using calculated large 3D reconstructed moiré superlattice, indicating that the moiré flat bands are heavily modulated by the 3D structural reconstruction. Further analysis shows that the Γ-point moiré potential is controlled by interlayer electron hybridization, while the K-point moiré potential is dominated by the strain redistribution of the moiré superlattice.

The schematic of our $WSe_2/WS_2$ heterostructure device is shown in Figure 1a. We used an array of graphene nanoribbons (GNRs) as contact electrodes, and the silicon substrate as a back gate to control the carrier density of the heterostructure. Details of the device fabrication are presented in Methods. Figure 1b shows an ambient AFM image of the top surface of the device: an array of GNRs (each separated by 100~200 nm) partially covers the $WSe_2/WS_2$ heterostructure. Figure 1c shows an enlarged large scale ultra-high vacuum (UHV) STM image of the heterostructure. The moiré superlattice can be clearly resolved in both the exposed TMD and GNR-covered areas, demonstrating the high quality of the heterostructure device.

Figure 2a shows a zoom-in STM image of the moiré superlattice in the exposed TMD area. It shows a moiré period of 8.16 nm, consistent with the period expected for an aligned $WSe_2/WS_2$ heterostructure with a near-zero twist angle. Importantly, the heterostructure shows large height variation at different sites within a moiré unit cell, resulting in an overall honeycomb lattice characterized by a large valley at each hexagon center that is surrounded by six peaks. The apparent height variation in an STM image always results from a convolution of geometric height change and electronic LDOS change, making it difficult to determine "true" height variation via STM. To eliminate this complication, we exploited the graphene covered TMD region, where the thin graphene layer covers the moiré superlattice conformally, but the electronic LDOS of graphene remains nearly constant. Figure 2b displays an STM topography image of the moiré superlattice in the graphene covered area using a bias of -0.19V. This bias lies within the TMD gap, and so all of the tunnel current flows through the graphene layer. The similarity between Figure 2b and Figure 2a confirms that the $WSe_2/WS_2$ topographical landscape does, in fact, feature six peaks surrounding one valley.

The moiré superlattice is formed by a periodic change in the layer stacking configurations between the top $WSe_2$ and the bottom $WS_2$ layers. Three high-symmetry stacking configurations are illustrated in Figure 2d, and denoted as the $B^{W/S}$, $B^{Se/W}$, and AA stackings, respectively. $B^{W/S}$ and $B^{Se/W}$ in this case are AB stackings. A common explanation for height modulation in TMD heterostructures is the stacking-dependent layer separation: the AA stacking has the largest interlayer spacing due to steric hindrance from the overlap of Se atoms in the $WSe_2$ layer and S atom in the $WS_2$ layer, while the $B^{Se/W}$ and $B^{W/S}$ stacking sites have smaller interlayer separations. To better understand the structural reconstruction of the moiré pattern, we first carried out a forcefield-based simulation of the superlattice. Figure 2c shows the resulting interlayer spacing distribution in a vertical-only relaxed, free-standing simulated $WS_2/WSe_2$ superlattice which indeed exhibits a peak at the AA stacking site. This stacking-dependent layer separation, however, does not explain our experimental data since it predicts a peak (AA) surrounded by six valleys ($B^{W/S}$ and $B^{Se/W}$), exactly the opposite of the STM images shown in Figures 2a and 2b.

To account for our experimental observation, we must consider additional 3D reconstruction of the moiré heterostructure. Recent studies have suggested that 3D reconstruction of TMD moiré superlattices may be significant[24], but there have been no *ab initio* simulations of this effect. Our *ab initio* simulations of the $WSe_2/WS_2$ heterostructure reveal a strong moiré superlattice reconstruction that includes both a large in-plane strain distribution and a prominent out-of-plane buckling. Figure 2e shows the calculated in-plane strain distribution within the WSe$_2$ layer of the heterostructure. The heterostructure tends to increase the area of the interlayer locked AB stacking regions due to its lower energy than AA stacking. As a result, the WSe$_2$ layer get locally compressed at the AB stacking regions due to its larger lattice constant compared with WS$_2$. The residual tensile strain localizes to the AA stacking region (Figure 2e). The WS$_2$, with smaller lattice constant, has the opposite strain distribution (Figure S7). To partially release this strain, the hetero-bilayer reconstructs in 3D by an in-phase buckling in the out-of-plane direction (Figure 2h,i). The simulated height distribution of the top WSe$_2$ layer (Figure 2f) perfectly reproduces our STM image (Figures 2a,b). Figure 2g shows a side view of the 3D reconstructed heterostructure from both experiment and theory. The buckling above the AA "valley" causes the AB sites to rise. It is noteworthy that the presence of a hBN substrate only slightly reduces the buckling effect (see more details in SI). The simulated line profile agrees well our experimental data (Figure 2g).

The moiré superlattice reconstruction has a profound impact on the electronic properties of the moiré flat bands. To observe this effect we used STS to probe the local electronic structure of the WSe$_2$/WS$_2$ heterostructure. Figure 3a displays the STS dI/dV spectra acquired at different moiré sites for -3 V < V$_{Bias}$ < 2V. Differences are seen in the spectra obtained at different moiré sites, consistent with previous studies performed on bilayer heterostructure grown on graphite[21, 22].

We first focus our analysis of the STS spectra on the moiré flat band closest to the valence band edge, where the effects of strongly correlated states have been observed previously[1, 4, 5]. A challenge in the STM study of TMD materials is how to distinguish electronic states arising from K or Γ points in the single-layer BZ[21, 22]. Here we utilize two distinct features of the K-point states to identify them. We first use the fact that, due to the much larger in-plane momentum of K-point states, their wavefunction decays faster outside the TMD layer (see details in SI). Figure 3b shows the height dependent dI/dV spectra measured at one of the AB sites (ultimately confirmed as $B^{Se/W}$ site) in the moiré pattern. Two prominent peaks are observed near $V_{bias}$ = -1.7 V and $V_{bias}$ = -1.5 V. The peak near -1.5 V exhibits a much stronger height dependence than the peak near -1.7V, suggesting that the peaks at -1.5 V and -1.7 V correspond to electronic states at the K and Γ points, respectively. We next use the two facts that the TMD K-point electron wavefunctions have large in-plane momentum, and are mainly contributed by W $d$ orbital with angular momentum m = ± 2, which will induce atomic-scale alternating constructive and destructive interference pattern as illustrated in Figure 3d (see details in SI). The high-resolution dI/dV mapping at -1.5 V shows pronounced dI/dV signal oscillation over atomic-scale distances that match the $WSe_2$ lattice (Figure 3e), while the dI/dV mapping at -1.7 V more smoothly varies (Figure 3f). This behavior confirms that the -1.5 V peak originates from K-point states at the valence band edge, whereas the -1.7 V peak originates from Γ-point states.

We performed larger-scale dI/dV mapping to directly visualize the localization of the flat bands in the real-space. Figures 3g and 3i show dI/dV maps obtained at the two peak energies -1.52 V (labeled K1) and -1.73 V (labeled Γ1). The local density of states (LDOS) for both of these flat band states are found to be strongly localized at the $B^{Se/W}$ site. Atomic-scale site dependence in the dI/dV signal for K-point states is again reflected in the K1 dI/dV mapping. Figures 3h and 3j show the dI/dV mapping at slightly lower energies. The LDOS distribution is seen to change dramatically and now shows LDOS minima where previously there were maxima at the $B^{Se/W}$ site.

To better determine the energy-dependent LDOS of the moiré flat bands, Figure 3k shows a density plot of dI/dV spectra for the bias range -1.72V < $V_{bias}$ < -1.42 V along the $B^{W/S}$-$B^{Se/W}$-AA direction, indicated by the yellow path marked in Figure 3g. Figure 3l shows the same plot, but over a different bias range: -1.86 V < $V_{bias}$ < -1.63V. Figure 3k shows a prominent moiré flat band at the valence band minimum that is strongly localized at the $B^{Se/W}$ site. This K-point moiré flat band is isolated from deeper moiré flat bands by a gap of ~50 meV. Figure 3c displays a high-resolution dI/dV spectrum at the $B^{Se/W}$ site, which shows that the moiré flat band has a width of only 10 mV (after deconvolution), thus setting an upper limit on the bandwidth of the moiré flat band. The 10 meV bandwidth is quite narrow, the occupied moiré miniband in twisted bilayer graphene by comparison has an experimental bandwidth of 10~40 meV[17-20]. The narrowness of the moiré flat band in combination with the strong long-range Coulomb interactions in 2D semiconductors[24], makes this TMD heterostructure an excellent platform to explore highly correlated quantum phenomena.

In addition to the valence band edge moiré flat band, Figure 3k shows that deeper moiré flat bands having different wavefunction spatial characteristics are present, but these bands are not well isolated. One such flat band can be tentatively identified near -1.6 V in Figure 3h. The dI/dV map of this flat band (labeled K2) shows a ring around the center of the $B^{Se/W}$ site and a weak plateau at the $B^{W/S}$ site. The ring-shaped electron wavefunction around the $B^{Se/W}$ site is reminiscent of the first excited states of a harmonic oscillator for a potential well centered on this site.

Figure 3l shows deeper flat bands mostly originating from the Γ point. The Γ point moiré flat band closest to the valence band edge lies at -1.73 V and is localized to the $B^{Se/W}$ site. At the slightly deeper energy of -1.78 V a new wavefunction distribution is seen and the LDOS now shows a minimum at the $B^{Se/W}$ site (Figure 3j). The ring-shaped electron wavefunction around the $B^{Se/W}$ site in this case is once again reminiscent of the first excited states of a harmonic oscillator (additional dI/dV mapping and position-dependent dI/dV spectra are included in the SI).

To interpret the observed moiré minibands, we performed large-scale density functional theory (DFT) calculations on the forcefield-reconstructed moiré superlattice. Figure 4a shows the calculated valence band structure for the $WSe_2/WS_2$ moiré superlattice in the mini-BZ. The valence band edge is set at E = 0. The bands closest to the valence band edge (0 ~ -0.18eV) are folded from the K-point, whereas the deeper bands below -0.18 eV have mixed origins from both the K-point and the Γ-point. Their origins can be distinguished by their LDOS distributions in the out-of-plane direction (see detailed discussion in the SI). We labeled four important energy ranges (E1-E4) in Figure 4a. The top-most valence band (within E1) has a bandwidth of only ~10 meV and is separated from the next band (within E2) by ~30 meV. The narrow, energetically isolated nature of the top-most band is in quantitative agreement with our experimental observations (Figure 3c). The deeper bands tend to mix with each other and are thus hard to distinguish experimentally. The top-most bands folded from the Γ-point is within E3, while the next set of deeper bands are within E4. The states in the gap between E3 and E4 are folded from the K-point.

The spatial distribution of the calculated charge density for the flat band states agrees well with the STS results. Figures 4b-e show the calculated LDOS distribution averaged over the four different energy ranges E1-E4, as well as averaged over the out-of-plane direction. Figure 4b shows the calculated LDOS for E1, mainly the top-most flat band from the K-point. The LDOS is observed to localize strongly in the $B^{Se/W}$ region of the moiré unit cell. This matches the experimental behavior quite well (Figure 3g), and, in fact, allows us to distinguish the $B^{Se/W}$ region from the $B^{S/W}$ region in our experimental images. As we move deeper into the valence bands (E2) our simulation reveals a delocalization of the K-point LDOS and the formation of a node at the center of the $B^{Se/W}$ region (Figure 4c). This closely matches our experimental observations (Figure 3h). The Γ-point LDOS behaves similarly, as seen by the comparison of theoretical Figures 4d,e with experimental Figures 3i,j. Once again we see the top-most Γ-point states (E3) showing strong localization in the $B^{Se/W}$ region, as well as increased delocalization accompanied by the formation of a node at slightly deeper energies (E4). The calculated energy separation between the top-most K and the Γ states ($180 \pm 20$ meV between E1 and E3) is also very similar to the experimental energy separation between the K1 and Γ1 states ($205 \pm 18$ meV). We thus observe very reasonable agreement between experiment and theory for both the flat band energy separation and the flat band spatial LDOS distribution.

It is conventionally believed that the interlayer interaction variation is the dominant effect in modulating the electronic structure of moiré superlattices[2]. However, our DFT calculation reveals that the flat bands at the K-point are mainly a result of the deformation of the monolayer instead of a hybridization-induced interlayer potential. Similar K-point flat band behavior can be reproduced by calculating the electronic structure of a puckered monolayer $WSe_2$ extracted from the relaxed moiré superlattice (Figures S10). The Γ-point flat bands, on the other hand, do arise from inhomogeneous interlayer hybridization within the moiré superlattice. Additional discussion regarding the origin of the flat band localization is presented in the SI.

In summary, we find that 3D moiré reconstruction dominates the low-energy moiré electronic structure, resulting in a narrow moiré flat band with 10 meV bandwidth at the valence band maximum in $WS_2/WSe_2$ heterostructures. Such quantitative understanding of the atomic and electronic structure within a moiré superlattice lays a solid foundation for the future control of novel, strongly correlated phenomena in TMD-based moiré heterostructures.

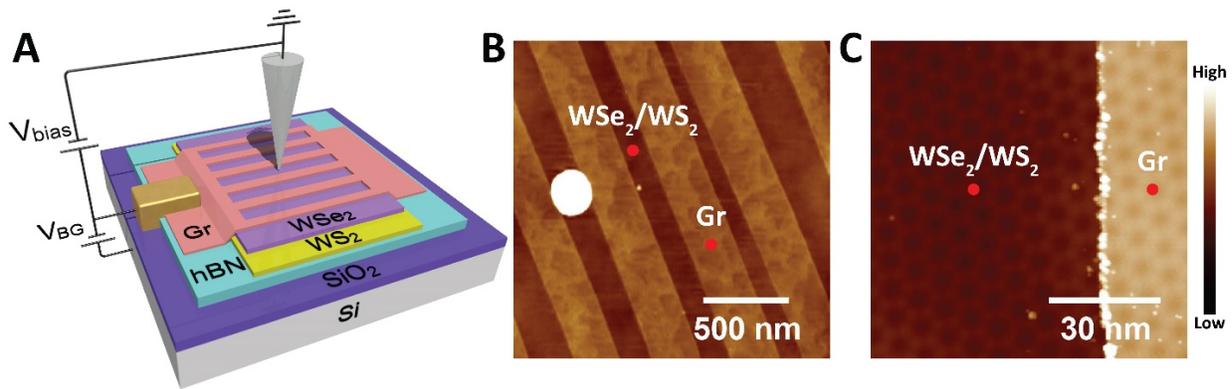

**Figure 1. Aligned WSe$_2$/WS$_2$ heterostructure. A**. Schematic of gate-tunable WSe$_2$/WS$_2$ heterostructure device used for STM study. Graphene nanoribbons (Gr) are placed on top of the WSe$_2$/WS$_2$ as contact electrodes. **B**. Room temperature ambient AFM image of the sample surface. Exposed WSe$_2$/WS$_2$ and graphene-covered areas are labeled. **C**. UHV STM image of the exposed WSe$_2$/WS$_2$ and graphene-covered area (T = 5.4K). A Moiré superlattice can be seen clearly in both areas. V$_{bias}$ = -3V, I = 100pA.

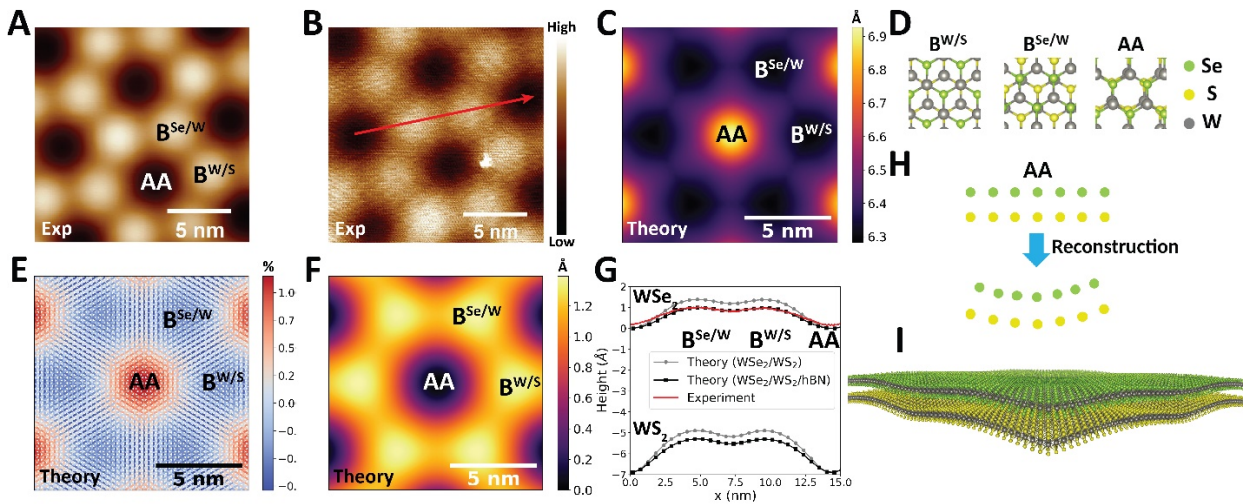

**Figure 2. Moiré superlattice reconstruction. A**. STM image of the exposed WSe$_2$/WS$_2$ area shows a moiré period of ~8nm. $V_{bias}$ = -3V, I = 100pA. **B**. STM image of the graphene-covered area ($V_{bias}$ = -0.19V. I = 100pA). The STM image here better reflects the true topography of the heterostructure (see text). **C**. Theoretical interlayer spacing distribution from simulation. **D**. Schematic of the three types of stacking: B$^{W/S}$, B$^{Se/W}$ and AA. **E**. Theoretical in-plane strain distribution (in %) for the WSe$_2$ layer from simulation. **F**. Theoretical height profile of the W atoms in the top WSe$_2$ layer from simulation. **G**. Calculated 3D buckling of the heterostructure and comparison to experiment. Black and gray dots show the simulated positions of W atoms for a freestanding heterostructure and a heterostructure supported by hBN, respectively. Red trace shows the experimental line-cut from B. **H**. Schematic of the buckling process. **I**. 3D view of the reconstructed WSe$_2$/WS$_2$ moiré superlattice from simulation.

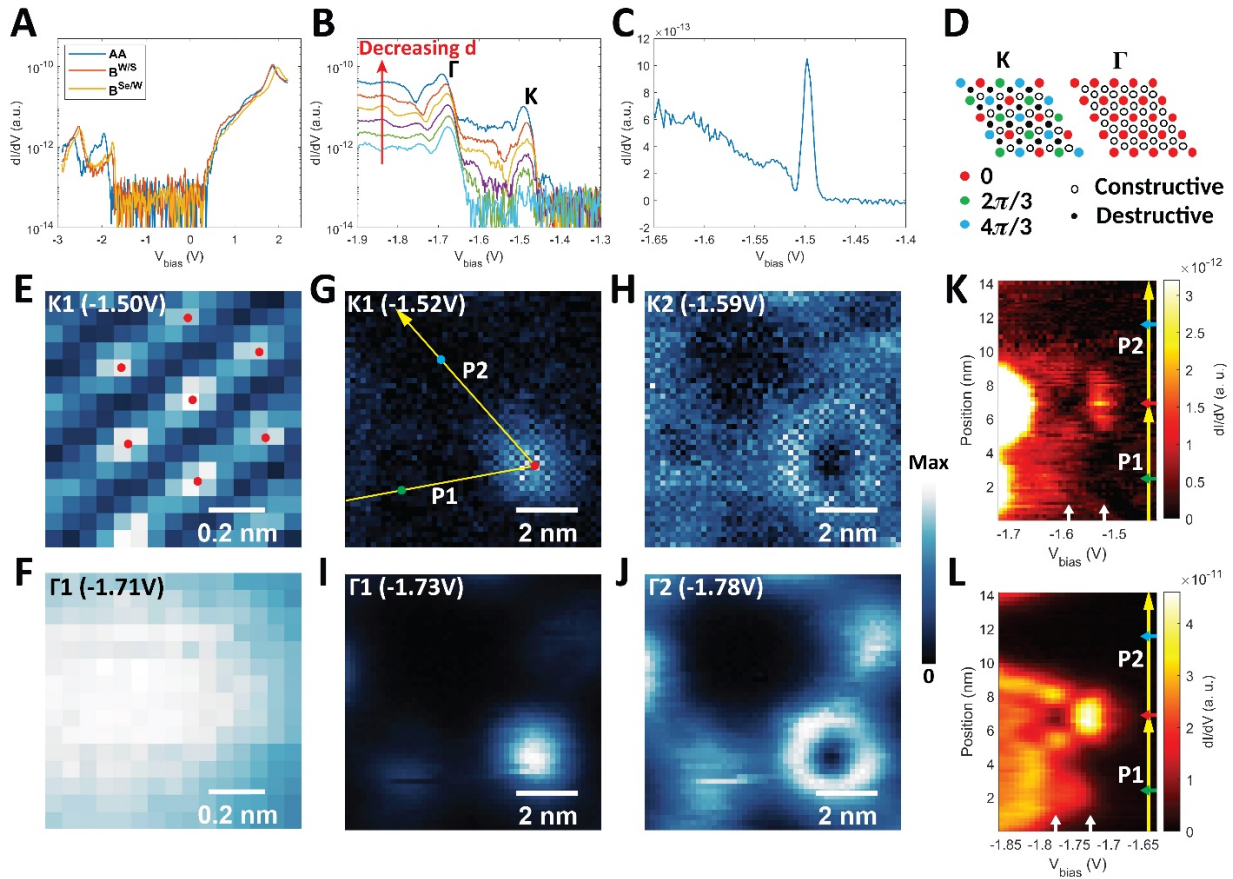

**Figure 3. Scanning tunneling spectroscopy (STS) measurement of moiré-induced flat bands. A**. Moiré-site dependent dI/dV spectra with low current setpoint ($V_{bias}$=-3V, I=70pA). Peaks in the -2V < $V_{bias}$ < -1.7V range show strong moiré site-dependent peak positions. **B**. Tip-sample distance (d) dependent STS at the B$^{Se/W}$ site ($V_{bias}$ = -2.15V, I=50, 100, 200, 400, 800, 1600pA). A second peak near $V_{bias}$ = -1.5V emerges with decreased d, indicating that it has a larger decay constant and originates from K-point states. **C**. High-resolution dI/dV spectrum measured at the B$^{Se/W}$ site. A sharp peak with ~10 mV width can be observed near $V_{bias}$ = -1.5 V

**D**. Illustration of the atomic-scale wavefunction interference pattern. K-point states have a 2π phase-winding over adjacent three W atoms, while Γ-point states have identical phases over all Se atoms sites. **E,F**. High resolution dI/dV mappings measured at the same $B^{Se/W}$ region with biases corresponding to (**E**) the K-point (-1.50V) and (**F**) the Γ-point (-1.71V) peaks. Red dots in **e** show the atomic lattice of the WSe$_2$. **G,H**. Large scale dI/dV mappings of K-point states for (**G**) $V_{bias}$ = -1.52V and (**H**) $V_{bias}$ = -1.59V. **I,J**. Large scale dI/dV mappings and Γ-point states for (**I**) $V_{bias}$ = -1.73V and (**J**) $V_{bias}$ = -1.78V. **G-J** show the same region of the sample surface. Solid dots in G label the positions of $B^{Se/W}$ (red), $B^{W/S}$ (green), and AA (blue) sites. **K,L**. dI/dV density plot of (**K**) K-point and (**L**) Γ-point states along the two-segment yellow path shown in G. Corresponding path is also shown in K,L. Horizontal arrows label the positions of the $B^{W/S}$ (green), $B^{Se/W}$ (red) and AA (blue) sites. White vertical arrows label the energies used in G-J. $V_{BG}$=50V for all measurement shown. The tip-sample distance is determined by the setpoint $V_{bias}$ = -2.15V, I=800pA for all spectroscopy shown. Lock-in modulation = 20mV for all panels except **c**, where the lock-in modulation is 5mV.

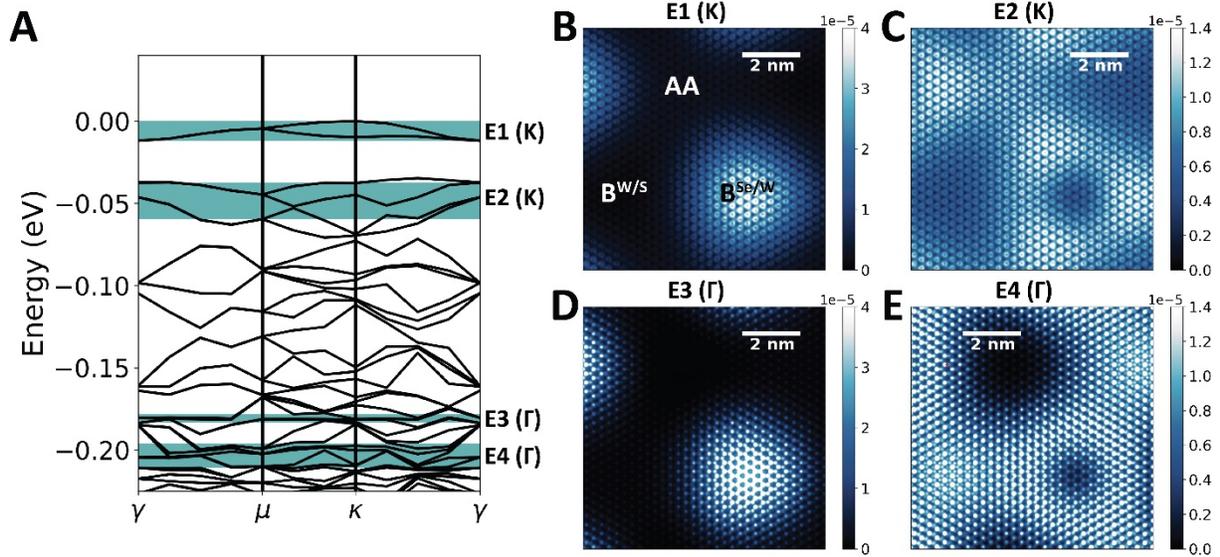

**Figure 4.** *Ab initio* **calculations of the electronic structure in reconstructed moiré superlattice. A**. Calculated electronic band structure plotted in the folded mini-BZ. Four important energy ranges (E1-E4) are labeled (green shaded areas) to highlight the top-most states folded from the K-point (E1, E2) and Γ-point (E3, E4). **B-E**. Calculated LDOS maps over a patch of area corresponding to the region used for dI/dV maps in Figures 3g-i. The LDOS maps are averaged over different energy ranges as labeled in panel A: **B**. E1, originating from K-point, **C**. E2, originating from K-point, **D**. E3, originating from Γ-point, and **E**. E4, originating from Γ-point. The maps are also averaged over the out-plane direction.

**Corresponding Author**

* Email: swli@berkeley.edu (S.L.), sglouie@berkeley.edu (S.G.L.), fengwang76@berkeley.edu (F.W.), and crommie@physics.berkeley.edu (M.C.)

**Author Contributions**

M.C., F.W. and S.L. conceived the project, and S.G.L. supervised the theoretical calculation. H.L., S.L. performed the STM measurement, M.H.N. carried out the DFT calculation. H.L., J.X., X.L. J.W., W.Z., S.Z. and S.K. fabricated the heterostructure device. E.R. and D.W. performed the SHG measurement. K.Y., M.B. and S.T. grew $WSe_2$ and $WS_2$ crystals. K.W. and T.T. grew the hBN single crystal. All authors discussed the results and wrote the manuscript.

**Notes**

The authors declare no financial competing interests.

# Supplementary Information for

# Imaging moiré flat bands in 3D reconstructed WSe$_2$/WS$_2$ superlattices

Hongyuan Li[1, 2, 3, 8], Shaowei Li[1, 3, 4, 8]*, Mit H. Naik[1, 3, 8], Jingxu Xie[1], Xinyu Li[1], Jiayin Wang[1], Emma Regan[1, 2, 3], Danqing Wang[1, 2], Wenyu Zhao[1], Sihan Zhao[1], Salman Kahn[1], Kentaro Yumigeta[5], Mark Blei[5], Takashi Taniguchi[6], Kenji Watanabe[7], Sefaattin Tongay[5], Alex Zettl[1, 3, 4], Steven G. Louie[1, 3]*, Feng Wang[1, 3, 4]*, Michael F. Crommie[1, 3, 4]*

[1]Department of Physics, University of California at Berkeley, Berkeley, CA, USA.

[2]Graduate Group in Applied Science and Technology, University of California at Berkeley, Berkeley, CA, USA.

[3]Materials Sciences Division, Lawrence Berkeley National Laboratory, Berkeley, CA, USA.

[4]Kavli Energy Nano Sciences Institute at the University of California Berkeley and the Lawrence Berkeley National Laboratory, Berkeley, CA, USA.

[5]School for Engineering of Matter, Transport and Energy, Arizona State University, Tempe, AZ, USA.

[6]International Center for Materials Nanoarchitectonics, National Institute for Materials Science, Tsukuba, Japan

[7]Research Center for Functional Materials, National Institute for Materials Science, Tsukuba, Japan

[8]These authors contributed equally: Hongyuan Li, Shaowei Li and Mit H. Naik

1. Sample fabrication
2. STM measurement
3. Determination of WSe$_2$ and WS$_2$ alignment via second-harmonic generation
4. Additional dI/dV mapping data
5. Computation methods
6. Determination of WS$_2$-hBN interlayer forcefield
7. Origin of 3D reconstruction in the moiré pattern
8. Identification of K-point and Γ-point peaks in STS measurement
9. Identification of K-point and Γ-point wavefunctions in simulation
10. Origins of the moiré for K-point and Γ-point flat bands

1. **Sample fabrication**

   The graphene/WSe$_2$/WS$_2$/hBN stack is fabricated using the micro-mechanical stacking techniques. A poly(propylene) carbonate (PPC) film stamp were used to pick up all exfoliated 2D material flakes in the following order: hBN, WS$_2$, WSe$_2$, and then graphene nanoribbons. The PPC film together with the stacked sample was then peeled, flipped over, and transfered on a SiO$_2$/Si substrate (SiO$_2$ thickness 285nm). The PPC layer is subsequently removed using ultrahigh vacuum annealing at 230 °C, resulting in an atomically-clean heterostructure suitable for STM measurements. The GNR array here provides a relatively low drain resistance for STM measurement of the TMD heterostructure on an insulating substrate, since graphene electrodes have lower contact resistance compared to deposited metal[1], and the STM tip can be placed very close to the graphene boundary. The graphene nanoribbon arrays were fabricated before the stacking process by cutting exfoliated graphene using the electrode-free local anodic oxidation technique[2]. 50nm Au and 5nm Cr were evaporated onto the graphene to make electric contact.

2. **STM measurement**

   STM and STS measurements were performed using a custom-built LT-STM (manufactured by Createc) at 5.4K in ultrahigh vacuum (pressure< $3 \times 10^{-10}$ Torr). Electrochemically etched Tungsten tips were cleaned via Argon ion sputtering and field emission. The lock-in frequencies used for the dI/dV measurement were between 600~900Hz.

3. **Determination of WSe$_2$ and WS$_2$ alignment via second-harmonic generation**

Polarization-dependent second harmonic generation (SHG) measurements were performed to determine the crystal axes of the $WS_2$ and $WSe_2$ layers[3, 4], and hence the twist angle between them. Figure S1 shows the polarization-dependent SHG signal on nonoverlapping monolayer $WSe_2$ (red dots) and $WS_2$ (black dots) regions, and the corresponding line-fits (red and black curves, respectively). The intensity of the SHG signal in the overlapping area is stronger than that measured in the nonoverlapping area, indicating a 0° twist angle instead of 60°.

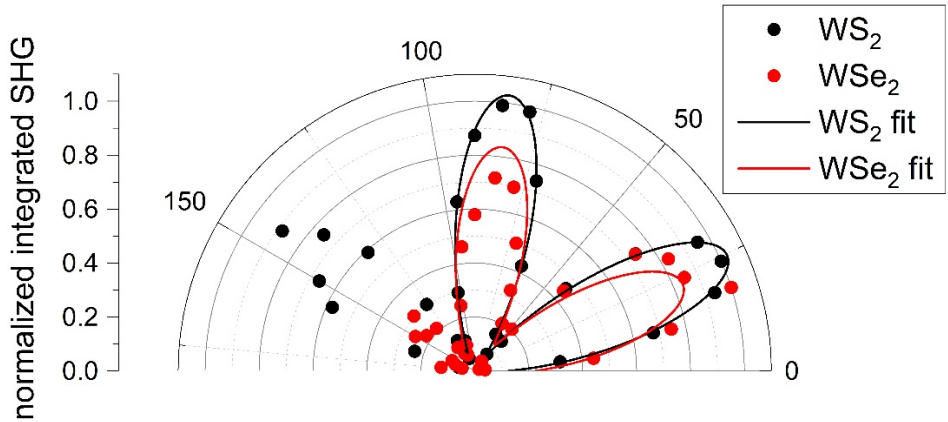

**Figure S1.** Polarization-dependent SHG signal on nonoverlapping monolayer $WS_2$ (black dots) and $WSe_2$ (red dots) regions and the corresponding fittings (black and red curves, respectively).

4. **Additional dI/dV mapping data**

Figures S2,S3 present additional data showing the voltage evolution of the dI/dV mappings. Figure S2a shows the topography of the area where the dI/dV mapping was performed. The colored dots indicate 3 different moiré sites. Figures S2b-S2t show the evolution of the dI/dV mapping with bias. Figures S2u and S2v show the same position-dependent dI/dV

spectra along the red lines in Figure S2d, but with different color bars, to highlight the K- and Γ-point minibands, respectively. Figure S3 reproduces the measurement in Figure S2 but with a different tip condition. Figure S4 shows a large scale dI/dV mapping of the lowest energy flat band obtained at $V_{bias}$ = -1.52V. For all the dI/dV mappings shown here the tip-sample distance was determined by following the setpoint: $V_{bias}$ = -2.15V, I = 800pA.

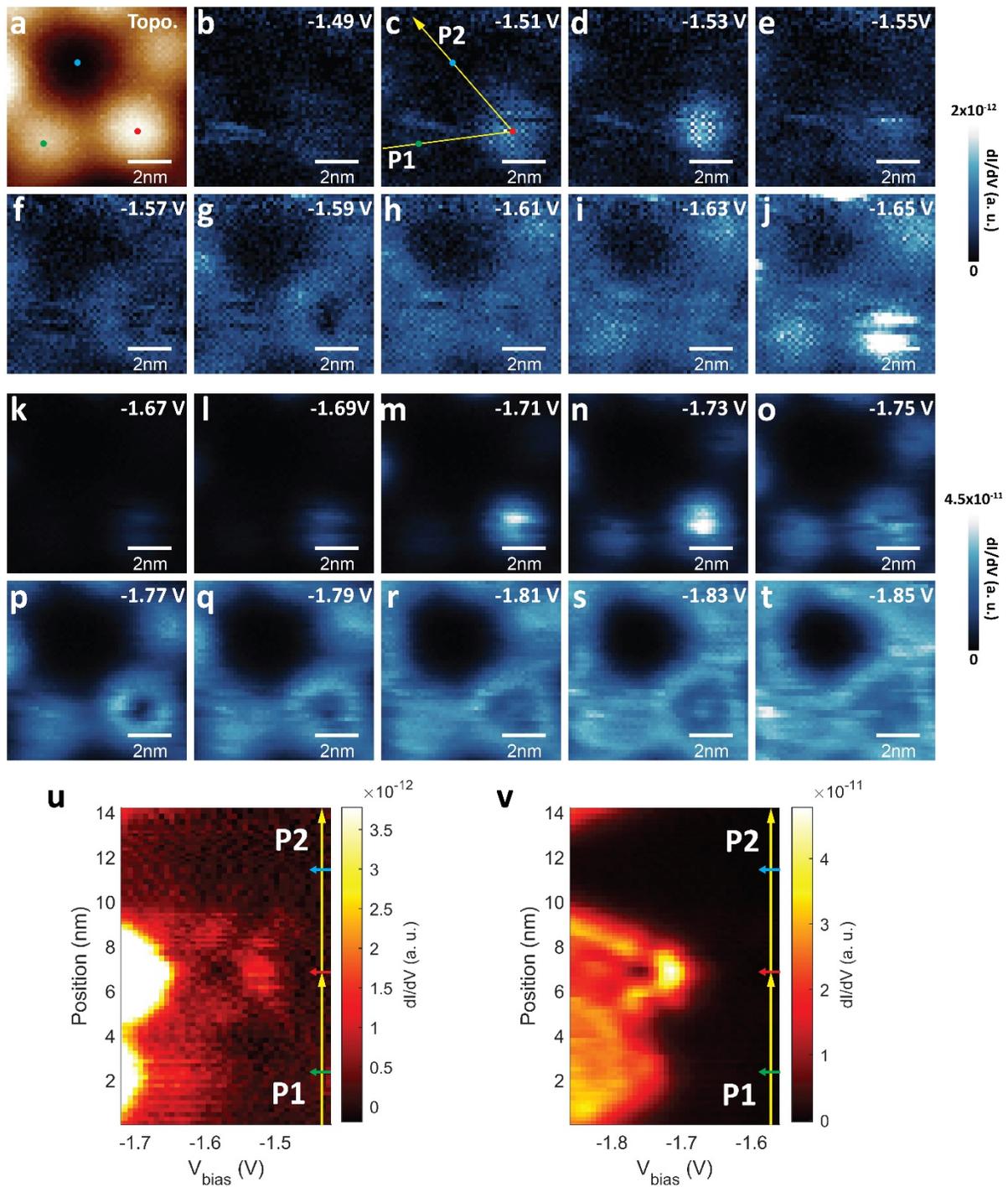

**Figure S2. Extra data #1: Voltage evolution of the dI/dV mappings. a.** Topography in the area used for dI/dV maps. Solid points label the positions of the $B^{Se/W}$ (red), $B^{W/S}$ (green), and AA (blue) moiré sites. **b-t**. Evolution of the dI/dV mappings corresponding to the (**b-j**) K-point

minibands and (**k-t**) Γ-point minibands. Solid points in **c** label the positions of the $B^{Se/W}$ (red), $B^{W/S}$ (green), and AA (blue) moiré sites. **u,v.** dI/dV density plot of (**u**) K-point and (**v**) Γ-point states along the two-segment yellow path shown in **c**. Corresponding path is also shown in **u,v**. Horizontal arrows label the positions of the $B^{W/S}$ (green), $B^{Se/W}$ (red) and AA (blue) sites. Tip-sample distance determined by the setpoint: $V_{bias}$ = -2.15V, I = 800pA.

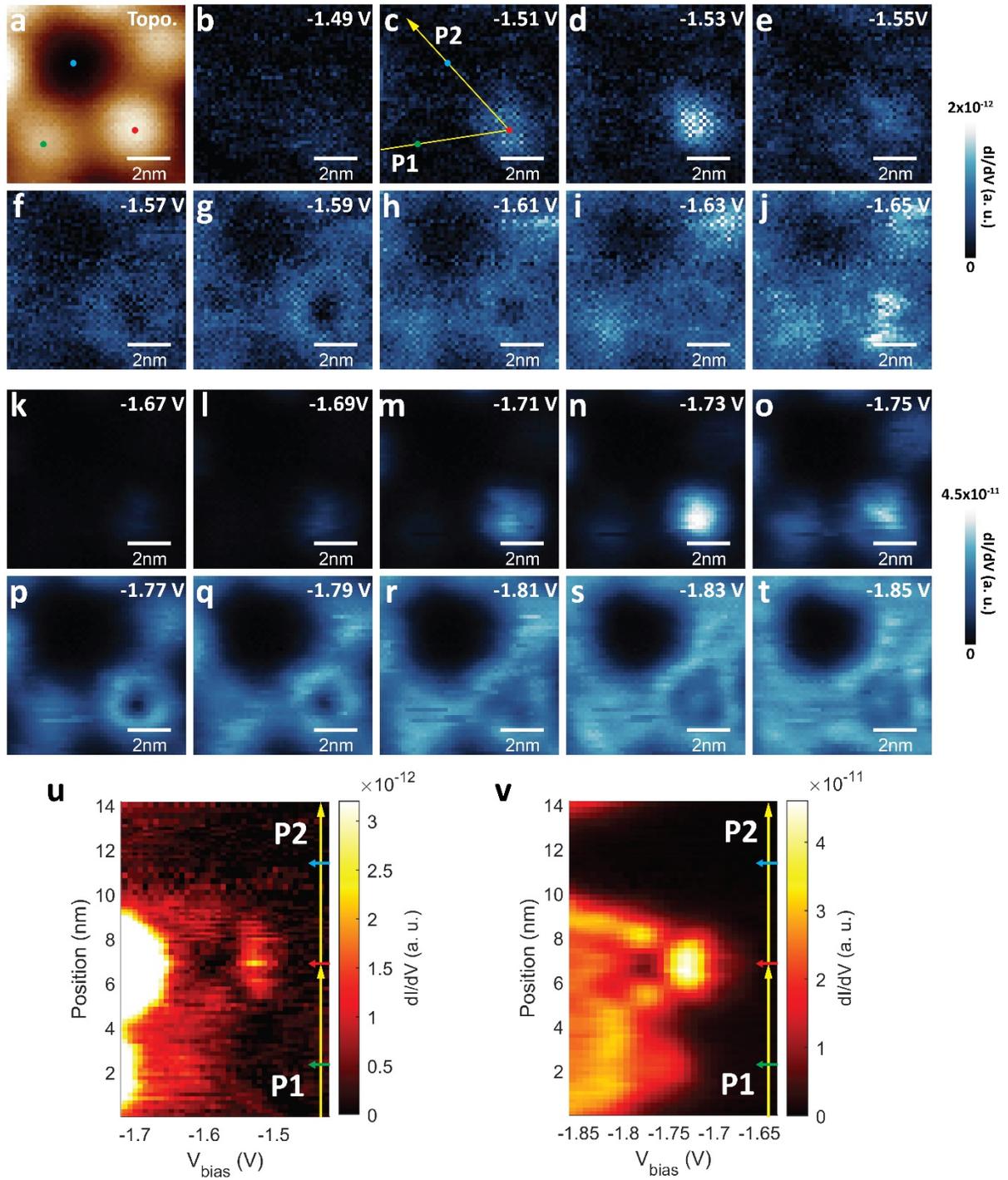

**Figure S3. Extra data #2: Voltage evolution of the dI/dV mappings. a**. Topography in the area used for dI/dV maps. Solid points label the positions of the $B^{Se/W}$ (red), $B^{W/S}$ (green), and AA (blue) moiré sites. **b-t**. Evolution of the dI/dV mappings corresponding to the (**b-j**) K-point

minibands and (**k-t**) Γ-point minibands. Solid points in **c** label the positions of the $B^{Se/W}$ (red), $B^{W/S}$ (green), and AA (blue) moiré sites. **u,v.** dI/dV density plot of (**u**) K-point and (**v**) Γ-point states along the two-segment yellow path shown in **c**. Corresponding path is also shown in **u,v**. Horizontal arrows label the positions of the $B^{W/S}$ (green), $B^{Se/W}$ (red) and AA (blue) sites. Tip-sample distance determined by the setpoint: $V_{bias}$ = -2.15V, I = 800pA.

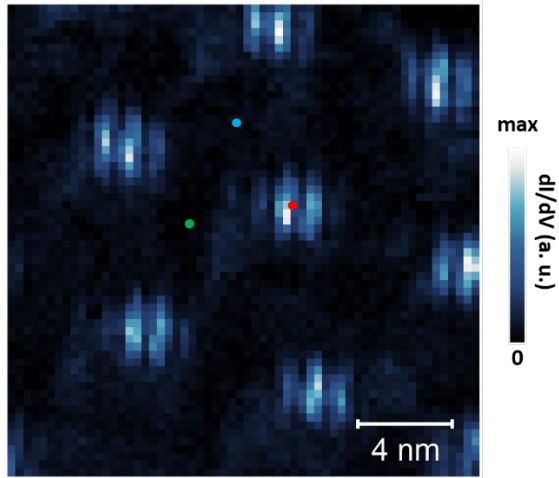

**Figure S4. Extra data #3: Large scale dI/dV mapping for top-most K-point flat band in valence manifold obtained at $V_{bias}$ = -1.52V.** Solid points label the positions of the $B^{Se/W}$ (red), $B^{W/S}$ (green), and AA (blue) moiré sites. Tip-sample distance was determined by the following setpoint: $V_{bias}$ = -2.15V, I = 800pA.

## 5. Computation methods

The simulated moiré superlattice was constructed using a 25x25 supercell of $WSe_2$ (lattice constant: 3.32 Å) and a 26x26 supercell of $WS_2$ (lattice constant: 3.19 Å). The mismatch between the supercell lattices of the two layers in this configuration is less than 0.1%. Structural relaxation of the moiré pattern was performed using the LAMMPS[5] package. The intralayer interactions were described using the Stillinger-Weber[6, 7] forcefield for the TMD layers and Tersoff[8] forcefield for hBN. The interlayer interactions were described using the recently parameterized Kolmogorov-Crespi (KC) interlayer potential[9, 10]. The KC potential has been shown to yield moiré structural reconstructions in good agreement with van der Waals corrected DFT calculations[10]. The force minimization was performed using the conjugate gradient method with a tolerance of $10^{-6}$ eV/Å. The simulation cell used in the forcefield reconstruction consisted of 8 moiré unit-cells. The moiré periodicity was maintained after structural relaxation. The electronic structure calculations (based on density functional theory[11] (DFT)) were performed on the reconstructed moiré unit-cell of 3903 atoms using the SIESTA[12] package. A supercell size of 21 Å was used in the out-of-plane direction to overcome spurious interactions between periodically repeated slabs. Spin-orbit coupling effects were included in the DFT calculation. The wavefunctions were expanded using a double-$\zeta$ plus polarization basis. The local density approximation[13] to the exchange-correlation functional was employed along with norm-conserving[14] pseudopotentials. A wavefunction cut-off of 80 Ry was used and only the $\Gamma$ point was sampled in the BZ to obtain the self-consistent charge density. Van der Waals corrections to DFT influence only the structural properties and not the DFT eigenvalues. Since we use structures obtained from forcefield reconstructions, we did not include van der Waals corrections in our DFT calculations.

The computation methods mentioned above are used for all the simulations in the main text and the SI except the Section. 8 of the SI.

## 6. Determination of WS$_2$-hBN interlayer forcefield

To account for the effect of the hBN substrate on the moiré superlattice in our simulation, we fit the Kolmogorov-Crespi[10] (KC) potential to our van der Waals corrected density functional theory (DFT) calculations. The DFT calculations were performed on a commensurate supercell of $WS_2$ and hBN to take into account their different lattice constants. We find that a 5x5 supercell of hBN matches a 4x4 supercell of $WS_2$. We consider only a single sheet of hBN in the simulation (Figure S5a). The binding energy of the $WS_2$/hBN supercell is computed for varying interlayer spacing as shown in Fig. S5b. The KC forcefield was fit to this data. We find that translating the $WS_2$ layer with respect to hBN does not affect the total energy (Figure S5c). The DFT calculations for fitting the forcefield are carried out using the Quantum Espresso[15] package with ultrasoft pseudopotentials[16] and the local density approximation[17] to the exchange-correlation functional. The wavefunctions and charge density are expanded in plane-waves up to an energy cut-off of 50 Ry and 500 Ry, respectively. The van der Waals interaction used in these calculations is the van der Waals density functional with Cooper exchange[18-20] (vdW-DF-C09). This same functional was used in the parameterization of the $WS_2$-$WSe_2$ interlayer interaction[9].

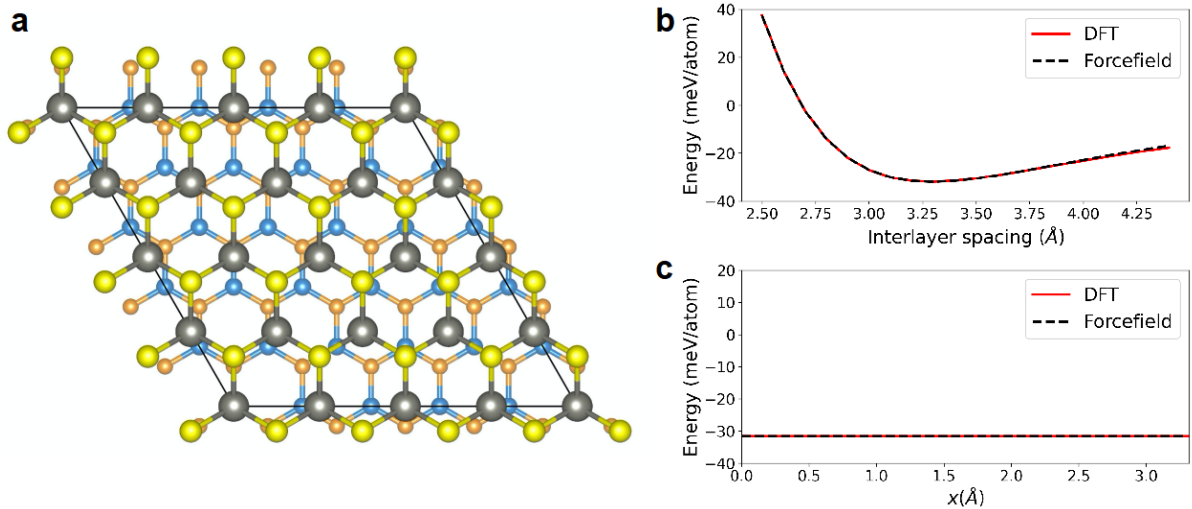

**Figure S5. Forcefield fitting of the WS$_2$-hBN interaction. a**. Commensurate supercell of WS$_2$/hBN heterostructure used in the simulation. **b**. Comparison of the binding energy as a function of interlayer spacing computed using van der Waals corrected DFT and forcefield. **c**. Comparison of binding energy as a function of displacement of the WS$_2$ layer with respect to the hBN layer at a fixed interlayer spacing of 3.2 Å.

7.  **Origin of 3D reconstruction in the moiré pattern**

As described in the main text, we find a 3D reconstruction of the moiré superlattice in the form of buckling of the two layers in the out-of-plane direction in addition to a varying interlayer spacing. The simulation of the suspended WSe$_2$/WS$_2$ yields a slightly larger buckling than the experiment which utilized an hBN substrate. The simulation performed in the presence of a single sheet of hBN leads to a reduction in the buckling which agrees well with the experiment (Figure S6). The hBN layer buckles together with the WS2 layer. If the hBN layer is deliberately confined in a flat configuration, the buckling no longer occurs in the fully relaxed TMD layers (Figure S6). The interlayer spacing distribution, on the other hand, is unaffected by this constraint. We also compare the strain build-up in the moiré superlattice for the abovementioned three configurations as shown in Figure S7. The strain in WS$_2$ in the AA stacking region and the domain boundary region is higher for the third configuration (Figure S7c) where buckling is constrained. Conversely, the buckling reduces strain in the TMD layers and thus lowers the total energy of the moiré superlattice.

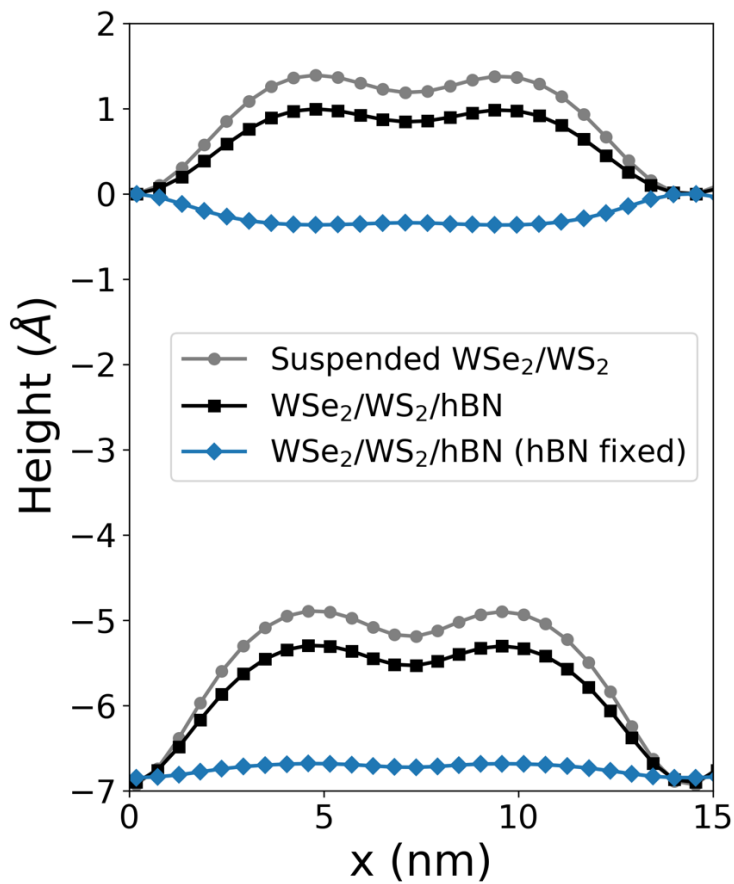

**Figure S6. Height of the TMD layer in different configurations from the simulation.** The buckling or in-phase bending is diminished slightly in the WSe$_2$/WS$_2$/hBN configuration when all layers are allowed to relax. When the hBN layer is fixed in a flat configuration, the buckling disappears. The interlayer spacing distribution is unchanged in the different configurations.

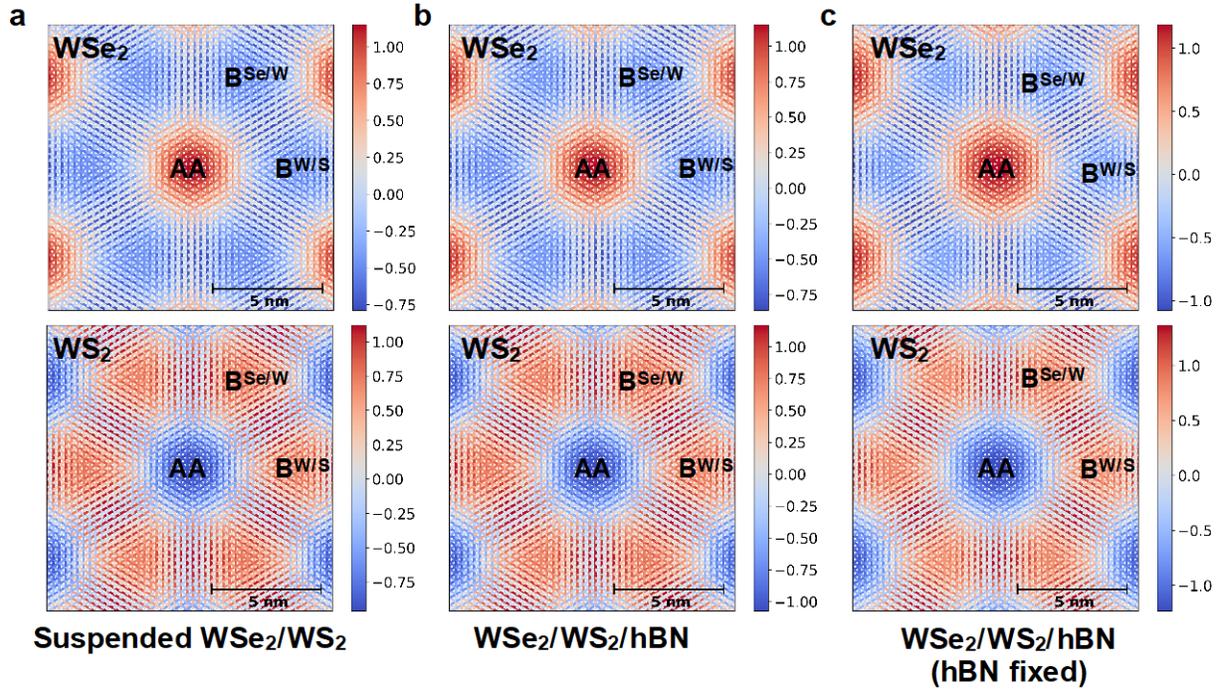

**Figure S7. Strain distribution in the WSe$_2$ and WS$_2$ layers for different configurations in the simulation. a**. suspended WSe$_2$/WS$_2$; **b**. WSe$_2$/WS$_2$/hBN; **c**. WSe$_2$/WS$_2$/hBN (fixed). The percentage strain is computed for each W-W nearest neighbor distance in the moiré. Color bar unit: %.

## 8. Identification of K-point and Γ-point peaks in STS measurement

Two distinct features of the K-point states were utilized to distinguish them from the Γ-point state. The first feature is that, due to the much larger in-plane momentum of K-point states, their wavefunction decays faster outside the TMD layer, as described by Eq.1:

$$k_\parallel^2 - \kappa^2 = \frac{2m_e(E-E_g)}{\hbar^2}, \qquad (1)$$

where $\mathbf{k}_\parallel$ is the in-plane wavevector, $\kappa$ is the decay constant in the out-of-plane direction, $m_e$ is the electron/hole effective mass, $E$ is the electron/hole energy, and $E_g$ is the energy of the band edge. This results in a much faster dI/dV signal decay rate with increased tip-sample distance for K-point states, as shown in Figure 3b of the main text.

The second feature is due to two facts. First, the electron wavefunctions at the K (K') point are mainly composed by the W $d$ orbital with angular momentum m = 2 (-2). As a result, these states gain a $4\pi/3$ phase when rotated by a $2\pi/3$ angle in space (-$4\pi/3$ phase gain for K' state). Second, due to the plane wave part $e^{i\mathbf{k}\cdot\mathbf{r}}$ of the Bloch wave, there will be a $2\pi$ phase winding for adjacent three W atoms (illustrated in Figure S8a). There phase factors, together, will induce atomic-scale alternating constructive and destructive interference pattern in the LDOS maps of the K(K')-point states. If we take the K-point state as an example, at the hollow circle points shown in the left panel in Figure S8a, the contribution of their nearby three adjacent W atoms will be,

$$1\cdot 1 + e^{\frac{2\pi i}{3}} \cdot e^{-\frac{4\pi i}{3}} + e^{\frac{4\pi i}{3}} \cdot e^{-\frac{8\pi i}{3}} = 0, \quad (2)$$

where the first factor in each term in the left-hand-side of the equation is the phase from the Bloch wave, while the second factor is from the angular momentum. Therefore, we will have destructive interference at the hollow circle points. For the solid circle points, the phase contribution from the Bloch wave keeps the same while the phase contribution from the angular momentum is reversed. The net effect will be,

$$1\cdot 1 + e^{\frac{2\pi i}{3}} \cdot e^{\frac{4\pi i}{3}} + e^{\frac{4\pi i}{3}} \cdot e^{\frac{8\pi i}{3}} = 3. \quad (3)$$

Therefore, we will have constructive interference at the solid circle points. For the K'-point state, both the wave vector of the Bloch wave and the angular momentum of the $d$ orbital are reversed, inducing that the final interference pattern is the same as the one of the K-point state. On the other hand, the Γ-point states are mainly contributed by the $p_z$ orital of the Se atoms, which has m = 0, and its Bloch wave has 0 wave vector. Therefore, all sites share the same phase and only have constructive interference (Figure S8a), which will make the LDOS spatial distribution of the Γ-point state smoother than the that of the K-point states.

Our analysis is confirmed by the DFT calculation results of a monolayer $WSe_2$. Here we simulate monolayer $WSe_2$ instead of the heterostructure since we believe the moiré potential works at a much larger length-scale (~8nm) and will have negligible impact on the atomic-scale (~0.3 nm) modulation. The LDOS maps in Figure 4 of the main text are generated by integrating out the wavefunctions in the out-of-plane direction, which are not exactly what the STS mappings measure. The STS mappings are basically the tunneling matrix element mapping. Assuming an s-wave tip-wavefunction, the tunneling matrix element can be estimated as the sample charge density at the s-wave center position[21]. To better simulate the STS mapping, we calculated the LDOS at various heights above the $WSe_2$ layer (3-8 Å, measured from the top Se atoms, which is the typical tip-sample distance in STM study), as shown in Figure S8b-S8i (Computation methods for this part is included at the end of this section). Figure S8b-S8e are the LDOS maps of the K-point states at different heights while Figure S8b-S8e are the LDOS maps of the Γ-point states. At low heights, both K-point and Γ-point maps show strong atomic-site dependent modulation due to the fact that the wavefunctions are mainly localized near the centers of the atomic sites. With height increasing, the K-point maps keep showing strong atomic-site dependent modulation, and the bright spots are located at one group of centers of three adjacent W atoms, while the other group of centers of three adjacent W atoms (Se sites) are dark. This result agrees well with our analysis of the K-point state wavefunction interference pattern. For the Γ-point maps, with height increasing, the maps become more and more uniform, which is consistent with our above analysis again.

Experimentally the real STM tip is usually non-ideal. Then the difference between the K-point and Γ-point mappings will manifest as the former will have strong spatial modulation at the atomic scale while the latter will be smoother.

**Computation method:**

The strictly-localized basis set we previously used to expand the moiré wavefunctions cannot precisely capture the decay of wavefunctions into vacuum[12, 22]. We hence perform the calculation on monolayer $WSe_2$ using a plane-wave basis expansion of the wavefunctions. The monolayer calculation is performed using the Quantum Espresso[15] package using ultrasoft[16] pseudopotentials and the local density approximation[17] to the exchange-correlation functional. The wavefunctions and charge density are expanded in plane-waves up to an energy cut-off of 50 Ry and 500 Ry, respectively. The Brillouin zone is sampled using a 12x12x1 Monkhorst-Pack7 grid. The charge density of the valence band maximum at the K and Γ point is plotted at a constant height above the $WSe_2$ layer (Figure S8b-S8i).

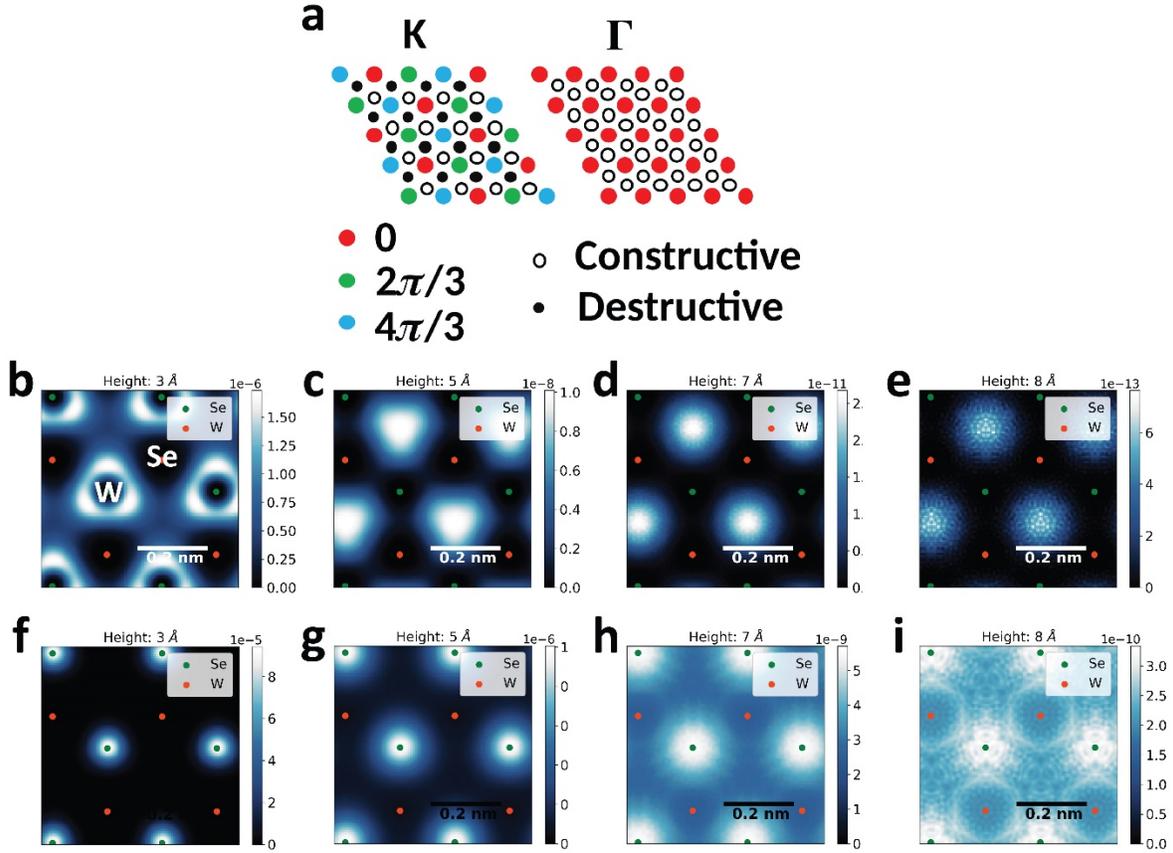

**Figure S8 Illustration of the atomic-scale wavefunction interference pattern. a**. K-point state has a 2π phase winding over adjacent three W atoms, while Γ-point state has identical phases over all Se atoms sites. **b-e**, Calculated LDOS mapping for the K-point state at different heights. **b**. 3 Å, **c**. 5 Å, **d**. 7 Å, and **e**. 8 Å. **f-i**, Calculated LDOS mapping for the K-point state at different heights. **f**. 3 Å, **g**. 5 Å, **h**. 7 Å, and **i**. 8 Å.

## 9. Identification of K-point and Γ-point wavefunctions in simulation

We identify the states in the moiré BZ as originating from the K or Γ point based on the localization of the wavefunction in the out-of-plane direction. Figure S9 shows the planar-averaged plot of the charge density of states arising from the K and Γ point. The K point states do not hybridize between the layers and are well localized in the WSe2 layer. The Γ point states on the other hand are strongly hybridized between the layers and hence have charge density on both layers.

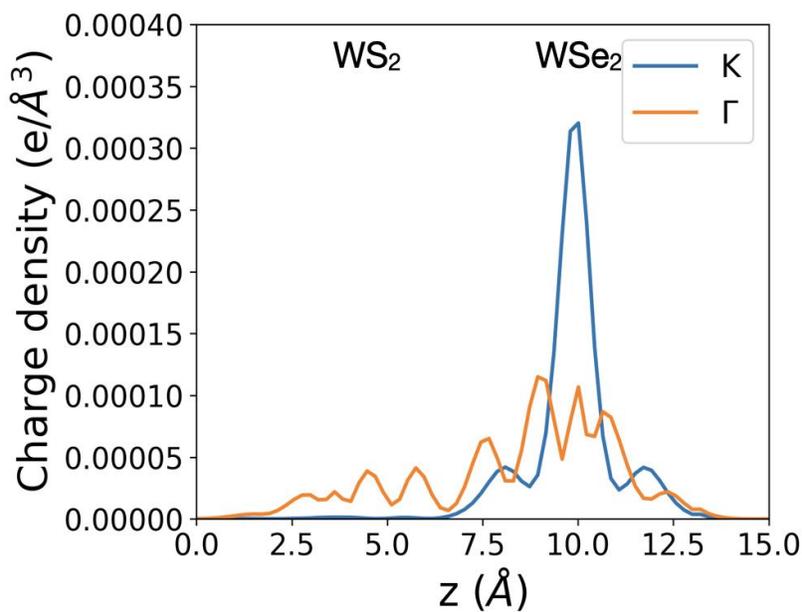

**Figure S9. Comparison of the charge density of states arising from the K and Γ points.** Planar-averaged charge density of states originating from the K (blue line) and Γ (orange line) point.

## 10. Origins of the moiré for K-point and Γ-point flat bands

As mentioned in the main text, the moiré potential for the K-point flat bands main originates from deformation of the WSe2 layer while the moiré potential for the Γ-point flat bands mainly originate inhomogeneous interlayer hybridization. The K-point wavefunctions, being of W-$d$ character, are highly localized within the WSe2 layer and almost do not hybridize in the bilayer, while the Γ-point wavefunctions, on the other hand, have a strong Se-$p_z$ character, leading to a stronger interlayer hybridization[23, 24]. The spatial varying interlayer spacing (larger at AA sites and smaller at AB sites) causes inhomogeneous interlayer hybridization for Γ-point states and thus the splitting of the Γ-point bands at different moiré sites[25, 26]. The origin of the K-point flat band moiré potential is more complicated. To understand this, we studied the electronic structure of an isolated puckered $WSe_2$ monolayer extracted from the relaxed hetero-bilayer moiré superlattice (Figure S10). The DFT calculations give very similar electronic band structures and spatial wavefunction distributions for both the relaxed hetero-bilayer and the puckered monolayer. That indicates that the moiré potential for the K-point flat bands thus originates mainly from the strains in the $WSe_2$ layer instead of the interlayer hybridization.

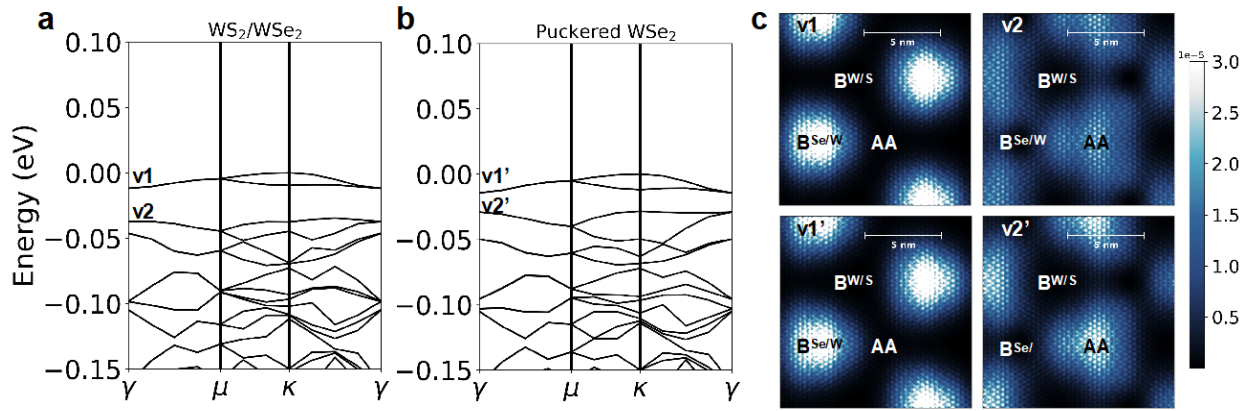

**Figure S10. Comparison of the electronic structure of a reconstructed $WS_2/WSe_2$ hetero-bilayer and an isolated puckered $WSe_2$ monolayer. a,b**. Comparison of the electronic band structures for (**a**) a relaxed $WSe_2/WS_2$ heterostructure and (**b**) an isolated puckered $WSe_2$ monolayer. The geometry of the puckered $WSe_2$ monolayer is extracted from the reconstructed hetero-bilayer moiré superlattice. **c**. Comparison of the LDOS spatial distribution of the $\gamma$-point states near the valence band edge marked in **a** and **b**. The LDOS maps are averaged over the out-plane direction.